\documentclass[11pt, a4paper]{article}

\newif\ifpublic\publictrue
\newif\iffancy\fancytrue

\usepackage{jheppub}
\usepackage{amsmath,amstext,amssymb,amscd}
\usepackage{graphicx}
\usepackage[usenames,dvipsnames]{xcolor}
\usepackage{bbm}
\title{Regge meets collinear in strongly-coupled $\mathcal{N}=4$ super Yang-Mills}
\author[a]{Martin Sprenger}
\affiliation[a]{Institut f\"ur Theoretische Physik, Eidgen\"ossische Technische Hochschule Z\"urich, Wolfgang-Pauli-Strasse 27, 8093 Z\"urich, Switzerland}
\emailAdd{sprengerm@itp.phys.ethz.ch}
\abstract{We revisit the calculation of the six-gluon remainder function in planar $\mathcal{N}=4$ super Yang-Mills theory from the strong coupling TBA in the multi-Regge limit and identify an infinite set of kinematically subleading terms. These new terms can be compared to the strong coupling limit of the finite-coupling expressions for the impact factor and the BFKL eigenvalue proposed by Basso et al.\ in \cite{Basso:2014pla}, which were obtained from an analytic continuation of the Wilson loop OPE. After comparing the results order by order in those subleading terms, we show that it is possible to precisely map both formalisms onto each other. A similar calculation can be carried out for the seven-gluon amplitude, the result of which shows that the central emission vertex does not become trivial at strong coupling.}
\keywords{Scattering amplitudes, AdS/CFT correspondence, Integrability}
\begin{document}

\newcommand{\ttt}{3\rightarrow 3}
\newcommand{\ttf}{2\rightarrow 4}
\newcommand{\Y}[2]{\mathrm{Y}_{#1}^{\left[#2\right]}}
\newcommand{\Yp}[2]{\mathrm{Y}_{#1}^{\prime\left[#2\right]}}
\newcommand{\U}[2]{\frac{\Y{2}{#1}{#2}}{1+\Y{2}{#1}{#2}}}
\newcommand{\Up}[2]{\frac{\Yp{2}{#1}{#2}}{1+\Yp{2}{#1}{#2}}}
\newcommand{\cratio}[4]{\frac{x_{#1,#2}^2 x_{#3,#4}^2}{x_{#3,#2}^2 x_{#1,#4}^2}}
\newcommand{\uh}{\hat{u}}
\newcommand{\vh}{\hat{v}}
\newcommand{\sqrtuhmg}{\left(\frac{\uh-1}{\uh+1}\right)^{\frac{1}{4}}}
\newcommand{\sqrtuhpg}{\left(\frac{\uh+1}{\uh-1}\right)^{\frac{1}{4}}}
\newcommand{\sqrtvhmg}{\left(\frac{\vh-1}{\vh+1}\right)^{\frac{1}{4}}}
\newcommand{\sqrtvhpg}{\left(\frac{\vh+1}{\vh-1}\right)^{\frac{1}{4}}}
\newcommand{\sqrtuhml}{\left(\frac{1-\uh}{1+\uh}\right)^{\frac{1}{4}}}
\newcommand{\sqrtuhpl}{\left(\frac{1+\uh}{1-\uh}\right)^{\frac{1}{4}}}
\newcommand{\sqrtvhml}{\left(\frac{1-\vh}{1+\vh}\right)^{\frac{1}{4}}}
\newcommand{\sqrtvhpl}{\left(\frac{1+\vh}{1-\vh}\right)^{\frac{1}{4}}}
\newcommand{\Yf}{\mathrm{Y}}

\setcounter{tocdepth}{2}
\maketitle

\section{Introduction}
\label{sec:introduction}

The high-energy behavior of scattering amplitudes in planar $\mathcal{N}=4$ super Yang-Mills (SYM) has been under active investigation recently.
A key feature of the high-energy limit, or more precisely multi-Regge limit, is that the perturbative expansion is naturally reorganized from an expansion in loops to an expansion in logarithmic orders.
Each of these logarithmic orders, starting from the leading logarithmic approximation (LLA) and followed by the (next-to-)$^k$-LLA (N$^k$LLA), is an approximation in kinematics, but resums contributions from all loop orders.
It is therefore an ideal tool to study the six-point remainder function in planar $\mathcal{N}=4$ SYM, for which a simple all-loop structure is expected.
While the full remainder function is now known up to five loops \cite{Dixon:2011pw, Dixon:2014voa, Dixon:2014iba, Dixon:2015iva, Caron-Huot:2016owq}, the step from a high number of loops to finite coupling still seems difficult.
One might therefore hope that understanding the all-loop structure in special kinematics, such as the high-energy limit or the collinear limit, gives further input that might help in unraveling the all-loop structure for full kinematics.\par
%%%%%%%%%%%%%
Indeed, in a series of papers it was shown that the six-point remainder function has a simple all-loop description in the multi-Regge limit, which takes the form of a dispersion relation-like integral \cite{Bartels:2008sc, Caron-Huot:2013fea}.
Physically, this integral describes a Regge cut contribution, which arises due to the formation of a bound state of two Reggeons.
The quantities determining the behavior of this bound state, called BFKL eigenvalue and impact factor, were determined in LLA \cite{Bartels:2008ce, Bartels:2008sc}, NLLA \cite{Lipatov:2010ad, Fadin:2011we}, N$^2$LLA \cite{Dixon:2012yy, Dixon:2013eka, Dixon:2014voa} and at strong coupling \cite{Bartels:2010ej, Bartels:2013dja}, before a finite-coupling proposal was put forward in \cite{Basso:2014pla}.\par
%%%%%%%%%%
Expanding the dispersion integral at weak coupling leads to an expansion of the remainder function in terms of the single-valued harmonic polylogarithms (SVHPLs) of \cite{Brown2004527}.
This observation allowed the generation of high-loop data \cite{Dixon:2012yy, Drummond:2015jea, Broedel:2015nfp}, but was also used to show that the remainder function in the multi-Regge limit has a simple all-loop structure even after carrying out the integration, at least in LLA \cite{Pennington:2012zj,Broedel:2015nfp}.
On the other hand, at strong coupling the dispersion integral should make contact with the multi-Regge limit of the semiclassical string result, which takes the form of a thermodynamic Bethe ansatz (TBA) \cite{Alday:2007hr, Alday:2009yn, Alday:2009dv, Alday:2010vh}.
This was first explored in \cite{Basso:2014pla}, where it was shown that the BFKL eigenvalue at strong coupling calculated from the dispersion integral agrees with the calculation from the TBA \cite{Bartels:2010ej, Bartels:2013dja}.
Obtaining a more detailed check at strong coupling is a key point of this paper.\par
%%%%%%%%%%%
For higher-point amplitudes, much less is known.
In the seven-point case, the interesting Mandelstam regions have been classified and the corresponding dispersion integrals in LLA have been constructed in \cite{Bartels:2013jna, Bartels:2014jya}, which in turn were evaluated up to five loops in \cite{Broedel:2016kls} for the MHV case.
Furthermore, the relevant Mandelstam regions have been investigated at strong coupling in \cite{Bartels:2014mka, Bartels:2014ppa}.
In the general $n$-gluon case, the multi-Regge limit of the symbol was investigated at two \cite{Bargheer:2015djt} and three \cite{Bargheer:2016eyp} loops in all Mandelstam regions and the generalization of the SVHPLs to the $n$-gluon setting were constructed in \cite{DelDuca:2016lad}, which allowed the authors to obtain the $n$-gluon MHV remainder function up to five loops in LLA.
From the point of view of Regge theory, all of the above calculations are still governed by a bound state of two Reggeons.
Therefore, the same BFKL eigenvalue and impact factor as in the six-point case appear.
However, there is one new ingredient in the seven-point case called the central emission vertex, which describes the emission of a physical gluon from a two-Reggeon bound state.
This central emission vertex is currently only known in LLA \cite{Bartels:2011ge}.
From the structure of the strong coupling result of \cite{Bartels:2014mka, Bartels:2014ppa}, one might conclude that the central emission vertex becomes trivial at strong coupling.
Showing that this is not the case is another point of this paper.\par
%%%%%%%%%%
Another special kinematic configuration is the collinear limit, which is governed by the Wilson loop OPE \cite{Alday:2010ku, Gaiotto:2010fk, Gaiotto:2011dt, Sever:2011da}.
This expansion takes the form of a flux-tube spanned by a light-like Wilson loop on which excitations propagate and interact.
The properties of these excitations, such as their dispersion relation, and the S-matrices describing how the excitations scatter are by now known at any value of the coupling constant \cite{Basso:2010in, Basso:2013vsa, Basso:2013aha, Belitsky:2014rba, Basso:2014koa, Basso:2014nra, Belitsky:2014sla, Belitsky:2014lta, Basso:2014hfa, Belitsky:2015efa, Basso:2015rta, Basso:2015uxa}.
Interestingly, it was shown recently that the contributions of some excitations can be resummed to obtain results in the multi-Regge limit \cite{Drummond:2015jea}, at strong coupling \cite{Fioravanti:2015dma, Bonini:2015lfr}, as well as the full amplitude for the tree-level NMHV case \cite{Cordova:2016woh} and the one-loop MHV case \cite{Lam:2016rel}. 
%%%%%%%%%%%
\par
Connections between the multi-Regge limit and the collinear limit were investigated perturbatively in \cite{Bartels:2011xy, Hatsuda:2014oza}.
However, the integrability of the flux-tube was only used in \cite{Basso:2014pla}, in which an analytic continuation connecting the two limits was used to propose finite-coupling expressions for the BFKL eigenvalue and the impact factor from finite-coupling expressions governing the energies and momenta of certain flux-tube excitations.
Remarkably, the leading term on the OPE side is sufficient to fully determine the quantities on the BFKL side.
The expansion for the multi-Regge limit obtained from the analytic continuation of the Wilson loop OPE will be referred to as BFKL OPE in the following.
These proposed finite-coupling expressions were so far checked against the available weak coupling data and the strong coupling result of \cite{Bartels:2010ej, Bartels:2013dja} and pass all tests.\par
In this paper, we identify kinematically subleading terms at strong coupling which allow a more detailed check of the finite-coupling expressions with the result of the TBA calculation.
After comparing those subleading pieces order by order, we will show that the two formalisms can in fact be exactly mapped onto each other.
This constitutes a strong check of the finite-coupling expressions put forward in \cite{Basso:2014pla}.
Furthermore, a similar calculation can be carried out for the seven-point amplitude, the result of which shows that the central emission vertex does not become trivial at strong coupling.
The result for the seven-point case furthermore provides predictions that could be checked against a potential finite-coupling expression for the central emission vertex.\par
%%%%%%%%%%%%%%%%%%%%
This paper is organized as follows.
In section \ref{sec:ttt_sc} we briefly review the calculation of the $\ttt$ -- amplitude at strong coupling from the TBA.
We then show how to obtain the subleading kinematic contributions in section \ref{sec:subleading_TBA} before explaining how to find those terms from the BFKL OPE in section \ref{sec:subleading_ope}.
Encouraged by the matching results, we show in section \ref{sec:matching} that the equations governing the TBA and the OPE result, respectively, can be mapped onto each other.
Finally, we examine a particular Mandelstam region of the $2\rightarrow 5$ -- amplitude in section \ref{sec:ttf} before concluding in section \ref{sec:conclusions}.
The technical details for the derivation of the impact factor from the BFKL OPE at strong coupling are presented in appendix \ref{sec:impact_factor}.
%%%%%%%%%%%%%%%%%%%%%%%%%%%%%%%%%%%%%%%%%%%%%%%%%%%%%%%%%%%%%%%%%%%%%%%%%%%%%%%

\section{The $3\rightarrow 3$ -- amplitude at strong coupling}
\label{sec:ttt_sc}
In this section, we briefly review the calculation of the $\ttt$ -- amplitude at strong coupling from the TBA.
Since this is simply an application of the algorithm developed in \cite{Bartels:2010ej, Bartels:2013dja, Bartels:2014ppa, Bartels:2014mka}, we will only present the pieces needed to follow the discussion of the new results.
The reader familiar with those references can immediately skip to section \ref{sec:subleading_TBA} for the new results.

\subsection{Six-point amplitude from the TBA}

As described in \cite{Alday:2009dv, Alday:2010vh} the six-point amplitude at strong coupling can be calculated as
\begin{equation}
	\mathcal{A}_6\sim e^{-\frac{\sqrt{\lambda}}{2\pi}A_{\mathrm{BDS}}+R_6},
	\label{eq:sc_amp}
\end{equation}
where $A_{\mathrm{BDS}}$ is the strong coupling extrapolation of the BDS-ansatz \cite{Bern:2005iz} and $R_6$ is the remainder function\footnote{Note that at strong coupling, there is no distinction between the MHV and the NMHV case. Differences only arise in contributions which are subleading in $\sqrt{\lambda}$.}, which depends only on the three dual-conformal cross ratios
\begin{equation}
	u_1=\cratio{2}{6}{3}{5},\quad u_2=\cratio{4}{6}{3}{1},\quad u_3=\cratio{2}{4}{1}{5},
	\label{eq:def_crs}
\end{equation}
where the dual variables $x_i$ are defined in terms of the gluon momenta $p_i$ as $p_i=:x_{i-1}-x_i$, with $x_{i+N}\equiv x_i$ and $x_{i,j}:=x_i-x_j$.
The amplitude is only fixed up to an overall normalization, because the prefactor in eq.~(\ref{eq:sc_amp}) is subleading in $\sqrt{\lambda}$.
The remainder function is given by several terms,	
\begin{equation}	
	R_6:=-\frac{\sqrt{\lambda}}{2\pi}\left(A_{\mathrm{free}}+A_{\mathrm{per}}+\Delta\right).
	\label{eq:def_r_sc}
\end{equation}
The simplest piece of eq.~(\ref{eq:def_r_sc}) is $\Delta$ which is directly given in terms of the cross ratios and reads
\begin{equation}
	\Delta = -\sum\limits_{i=1}^3\left(\frac{1}{8}\log^2 u_i+\frac{1}{4}\mathrm{Li}_2(1-u_i)\right).
	\label{eq:def_delta}
\end{equation}
To describe the other two pieces, we introduce three functions $\widetilde{\Yf}_a(\theta)$ with $a=\{1,2,3\}$, which depend on a spectral parameter $\theta$, as well as on the parameters $m=|m|e^{i\varphi}$ and $C$, which are auxiliary parameters that describe the kinematics, as we will see momentarily.
Those $\widetilde{\Yf}_a$-functions satisfy a set of TBA-like equations,
\begin{align}
	\log\widetilde{\Yf}_a(\theta)&=-m_a\cosh\theta-C_a-\sum\limits_{a'}\int\limits_{\mathbb{R}}d\theta' K_{aa'}(\theta-\theta')\log\left(1+\widetilde{\Yf}_{a'}(\theta')\right),
	\label{eq:def_Y_sys}
\end{align}
where the parameters are given by
\begin{equation}
	m_1=m_3=|m|,\,\, m_2=\sqrt{2}|m|,\quad C_1=-C_3=C, \,\, C_2=0
\end{equation}
and the integration kernels read 
\begin{equation}
	K_{aa'}(\theta)=\begin{pmatrix} K_1(\theta) & K_2(\theta) & K_1(\theta) \\ K_2(\theta) & 2 K_1(\theta) & K_2(\theta) \\ K_1(\theta) & K_2(\theta) & K_1(\theta) \end{pmatrix}
\end{equation}
in terms of the two functions
\begin{equation}
	K_1(\theta) = \frac{1}{2\pi}\frac{1}{\cosh\theta},\quad K_2(\theta) = \frac{\sqrt{2}}{\pi}\frac{\cosh\theta}{\cosh 2\theta}.
	\label{eq:def_kernels}
\end{equation}
In terms of these functions and the TBA parameters $m$ and $C$ the remaining contributions to the remainder function are given by
\begin{equation}
\begin{aligned}
	A_{\mathrm{free}}&=\frac{|m|}{2\pi}\int\limits_{\mathbb{R}}d\theta\cosh\theta\log\left[\left(1+\widetilde{\Yf}_1(\theta)\right)\left(1+\widetilde{\Yf}_3(\theta)\right)\left(1+\widetilde{\Yf}_2(\theta)\right)^{\sqrt{2}}\right]\,\mathrm{and}\\
	A_\mathrm{per}&=\frac{1}{4}|m|^2.
	\label{eq:def_afree}
\end{aligned}
\end{equation}
The cross ratios (\ref{eq:def_crs}) can be calculated from the $\widetilde{\Yf}_a$-functions through the relations
\begin{equation}
	u_1=\frac{\widetilde{\Yf}^{[-3]}_2(-i\varphi)}{1+\widetilde{\Yf}^{[-3]}_2(-i\varphi)},\quad u_2=\frac{\widetilde{\Yf}^{[-1]}_2(-i\varphi)}{1+\widetilde{\Yf}^{[-1]}_2(-i\varphi)},\quad u_3=\frac{\widetilde{\Yf}^{[1]}_2(-i\varphi)}{1+\widetilde{\Yf}^{[1]}_2(-i\varphi)},
	\label{eq:crs_yfcts}
\end{equation}
where $\widetilde{\Yf}_a^{[k]}(\theta):=\widetilde{\Yf}_a(\theta+i k\frac{\pi}{4})$, 
which provide the link between the TBA parameters and the kinematics of the scattering process.
Note that eq.~(\ref{eq:crs_yfcts}) is the only point in which a dependence on the parameter $\varphi$ arises.
For completeness, let us mention that the $\widetilde{\Yf}_a$-functions satisfy a recursion relation,
\begin{equation}
	\widetilde{\Yf}_a(\theta)=\frac{1}{\widetilde{\Yf}^{[2]}_{4-a}(\theta)\left(1+\frac{1}{\widetilde{\Yf}^{[1]}_{a+1}(\theta)}\right)\left(1+\frac{1}{\widetilde{\Yf}^{[1]}_{a-1}(\theta)}\right)}.
	\label{eq:def_recrel}
\end{equation}
This relation allows us to easily construct the $\widetilde{\Yf}_a$-functions far away from the real axis, where the integral representation (\ref{eq:def_Y_sys}) is tricky because of singularities of the integration kernels.\par
All of the above has a nice generalization to the general $n$-gluon case, but the expressions become more complex.
We therefore refer the reader to \cite{Alday:2010vh} for details.

\subsection{Multi-Regge kinematics}
\label{sec:r_eucl}
So far, we have described the TBA in general kinematics.
We now specialize to the multi-Regge limit of the $\ttt$ -- amplitude.
As described in \cite{Bartels:2010tx}, this limit is characterized by the following behavior of the cross ratios,
\begin{equation}
	u_1\rightarrow 1^+,\quad u_2\rightarrow 0^+,\quad u_3\rightarrow 0^+,
	\label{eq:mrl_crs}
\end{equation}
where the superscript means taking the limit from above.
This limit is taken such that the reduced cross ratios
\begin{equation}
	\tilde{u}_2:=\frac{u_2}{u_1-1},\quad \tilde{u}_3:=\frac{u_3}{u_1-1}
	\label{eq:red_crs_mrl}
\end{equation}
remain constant.
This differs from the $\ttf$ -- amplitude only in that the large cross ratio $u_1$ is now slightly larger than one, not smaller, see e.g.\ \cite{Bartels:2012gq}.
This entails some changes in the description of the kinematics in terms of the TBA parameters.
To see this, we start from the exact relation
\begin{equation}
	C=\cosh^{-1}\left(\frac{1-u_1-u_2-u_3}{2\sqrt{u_1u_2u_3}}\right)=\cosh^{-1}\left(-\frac{1+\tilde{u}_2+\tilde{u}_3}{2\sqrt{u_1\tilde{u}_2\tilde{u}_3}}\right),
	\label{eq:c_crs}
\end{equation}
which can be derived from the recursion relation (\ref{eq:def_recrel}) as well as the exact relation
\begin{equation}
	\frac{\widetilde{\Yf}_3(\theta)}{\widetilde{\Yf}_1(\theta)}=e^{2C},
\end{equation}
see eq.~(\ref{eq:def_Y_sys}).
For the limiting behavior (\ref{eq:mrl_crs}), the argument in eq.~(\ref{eq:c_crs}) is real and smaller than minus one, which leads to
\begin{equation}
	C=i\pi+\widetilde{C},
	\label{eq:def_x}
\end{equation}
with $\widetilde{C}$ being real.
For comparison, in the $\ttf$ -- case $C$ is purely imaginary.
This, however, is the only difference between the two kinematical settings.
We can therefore still follow the analysis of \cite{Bartels:2012gq} for the $\ttf$ -- case, and use the relations between the $\widetilde{\Yf}_a$-functions and the cross ratios (\ref{eq:crs_yfcts}) to see how the TBA parameters $m$ and $C$ behave in the multi-Regge limit.
The result is that the limit
\begin{equation}
	|m|\rightarrow\infty,\quad \varphi\rightarrow 0,\quad C\,\,\mathrm{const.}
	\label{eq:params_mphi}
\end{equation}
describes the multi-Regge regime.
In terms of the parameters $\varepsilon=e^{-|m| \cos\varphi}, w=e^{|m|\sin\varphi}$ which behave as $\varepsilon\rightarrow 0$ and $w\rightarrow\,\mathrm{const.}$ in the multi-Regge limit, we find the following parametrization of the cross ratios in terms of the TBA parameters
\begin{equation}
	u_1=1-\varepsilon\left(w+\frac{1}{w}+2\,\cosh C\right),\quad u_2=\varepsilon w,\quad u_3=\frac{\varepsilon}{w},
	\label{eq:param_tba}
\end{equation}
with corrections of $\mathcal{O}(\varepsilon^2)$.
This parametrization nicely shows the behavior (\ref{eq:mrl_crs}) once we take into account eq.~(\ref{eq:def_x}).

\subsection{Calculation of the remainder function $R_{\ttt}$}
\label{sec:calc_gen}
Having discussed the kinematics of the multi-Regge limit, we now turn to the evaluation of the remainder function (\ref{eq:def_r_sc}).
The two contributions $\Delta$ and $A_{\mathrm{per}}$ are easily computed.
Simply plugging in the parametrization (\ref{eq:param_tba}) and expanding in $\varepsilon$ we obtain
\begin{equation}
	\begin{aligned}
	\Delta &= -\frac{1}{4}\log^2\varepsilon-\frac{1}{4}\log^2 w-\frac{\pi^2}{12}+\mathcal{O}\left(\varepsilon\log\varepsilon\right),\\
	A_{\mathrm{per}} &= \frac{1}{4}\log^2\varepsilon + \frac{1}{4}\log^2 w.
	\end{aligned}
	\label{eq:eucl_contr}
\end{equation}
The contribution $A_{\mathrm{free}}$ is slightly more involved.
However, as explained in detail in \cite{Bartels:2012gq}, the limit (\ref{eq:params_mphi}) is special in that the integrals in both the TBA equations (\ref{eq:def_Y_sys}) and $A_{\mathrm{free}}$ (\ref{eq:def_afree}) are exponentially suppressed in $|m|$.
Indeed, a careful analysis shows that
\begin{equation}
	A_{\mathrm{free}} = \mathcal{O}(\varepsilon\log\varepsilon)
	\label{}
\end{equation}
and is therefore negligible in the limit $\varepsilon\rightarrow 0$, see \cite{Bartels:2010ej} for details.
Summing up all contributions, we find that the remainder function is a constant
\begin{equation}
	R_{\ttt}=\frac{\sqrt{\lambda}}{2\pi}\frac{\pi^2}{12}.
\end{equation}
This constant, however, comes solely from the $\mathrm{Li}_2$-part of $\Delta$ and cancels with a similar term in $A_{\mathrm{BDS}}$.
We therefore conclude that the remainder function is trivial in the limit (\ref{eq:params_mphi}).
This is in accordance with the weak coupling result \cite{Bartels:2010tx} and can be traced back to the absence of Regge cut contributions in this kinematic region.
However, as shown in \cite{Bartels:2010tx}, there is a kinematic region of the $\ttt$ -- amplitude, in which a Regge cut is known to appear.
This so-called Mandelstam region can be probed by first performing an analytic continuation in the cross ratios as
\begin{equation}
	u_1\rightarrow e^{2i\alpha} u_1,\quad  u_2\rightarrow e^{i\alpha}u_2,\quad  u_3\rightarrow e^{i\alpha}u_3,\,\mathrm{with}\,\alpha= 0\dots\pi,
	\label{eq:def_regge_region}
\end{equation}
and only then taking the multi-Regge limit.
In this kinematic region we therefore expect to find a non-trivial remainder function.

\subsubsection{$R_{\ttt}$ in the Mandelstam region}

Probing the remainder function at strong coupling in different Mandelstam regions was explored in detail in \cite{Bartels:2010ej, Bartels:2014ppa, Bartels:2014mka}, where an algorithm for the analytic continuation of a general $n$-gluon amplitude is presented.
In the following, we just present the key concepts and refer the reader to those references for details.\par
As explained in the last section, we want to perform an analytic continuation in the cross ratios.
This is trivial for the $\Delta$-contribution to the remainder function (\ref{eq:def_delta}), as it is already expressed in cross ratios, but it is more involved for the contributions $A_\mathrm{free}$ and $A_\mathrm{per}$.
Looking at the TBA equations (\ref{eq:def_Y_sys}) and how they are related to the cross ratios (\ref{eq:crs_yfcts}), it is clear that an analytic continuation in the cross ratios corresponds to an analytic continuation in the TBA parameters.
Such a continuation in the TBA parameters is subtle for the following reason.
For any given $\widetilde{\Yf}_a$-function there are locations in the complex $\theta$-plane where $\widetilde{\Yf}_a(\theta)=-1$.
The location of these points, of course, depends on the TBA parameters.
Hence these points will move in the complex $\theta$-plane during the analytic continuation.
However, these are very special points from the point of view of the TBA, as they correspond to poles of the integrands in eq.~(\ref{eq:def_Y_sys}).
Therefore, if any of those points approach the integration contour during the analytic continuation, we have to deform the contour such that we avoid having a pole on the line of integration.
At the end of the continuation, we want to compare the result with the original equations, so we have to pull back the integration contour to the real axis.
In doing so, we will hit those poles which have crossed the real axis, in which case we have to pick up residue contributions.
We parametrize the location of those poles by $\widetilde{\theta}_{a,i}$ with $i=1,\dots,n_a$, indicating 
which $\widetilde{\Yf}_a$-function they are associated to, i.e.\ we have
\begin{equation}
	\widetilde{\Yf}_a(\widetilde{\theta}_{a,i})=-1\quad\mathrm{for}\quad i=1,\dots,n_a.
	\label{}
\end{equation}
Picking up the residue contribution of those poles leads to a modification of the original TBA equations as
\begin{equation}
\begin{aligned}
	\log\widetilde{\Yf}'_a(\theta)=&-m'_a\cosh\theta-C'_a-\sum\limits_{a'}\int\limits_{\mathbb{R}}d\theta'K_{aa'}(\theta-\theta')\log\left(1+\widetilde{\Yf}'_{a'}(\theta')\right)\\
	&\quad-\sum\limits_{a'}\sum\limits_{i=1}^{n_a} \mathrm{sign}(\mathrm{Im}(\widetilde{\theta}_{a,i}))\log S_{aa'}(\theta-\widetilde{\theta}_{a,i}),
\end{aligned}
	\label{eq:mod_TBA}
\end{equation}
where we indicate the TBA parameters at the endpoint of the continuation with a prime.
The quantities $S_{aa'}(\theta)$ appearing in eq.~(\ref{eq:mod_TBA}) are related to the kernels via
\begin{equation}
	K_{aa'}(\theta)=:-\frac{1}{2\pi i}\partial_\theta\log S_{aa'}(\theta).
	\label{eq:def_smat}
\end{equation}
For the basic kernels $K_1(\theta)$ and $K_2(\theta)$ they explicitly read
\begin{equation}
	S_1(\theta)=i\frac{1-ie^{\theta}}{1+ie^{\theta}},\quad S_2(\theta)=\frac{2i\sinh\theta-\sqrt{2}}{2i\sinh\theta+\sqrt{2}}.
	\label{def_basic_smat}
\end{equation}
In terms of the analytically continued parameters we can then calculate the remaining contributions to the remainder function using
\begin{align}
	A_\mathrm{per}'=&\,\frac{1}{4}{|m|'}^2,\\
	A_\mathrm{free}'=&\,\frac{|m|'}{2\pi}\int\limits_{\mathbb{R}}d\theta\cosh\theta\log\left[\left(1+\widetilde{\Yf}'_1(\theta)\right)\left(1+\widetilde{\Yf}'_3(\theta)\right)\left(1+\widetilde{\Yf}'_2(\theta)\right)^{\sqrt{2}}\right]\label{eq:mod_a_free_per}\\
	&\quad+i|m|'\sum\limits_{a}\sum\limits_{i=1}^{n_a}\mathrm{sign}(\mathrm{Im}(\widetilde{\theta}_{a,i}))\sinh\widetilde{\theta}_{a,i}.\nonumber
\end{align}
We now clearly see the effect that the analytic continuation has had in the appearance of the residue terms in both the TBA equations and the $A'_{\mathrm{free}}$ -- contribution.
To obtain an explicit result for the $\ttt$ -- remainder function for the continuation (\ref{eq:def_regge_region}), all we need to do is figure out how many crossing solutions there are for the three $\widetilde{\Yf}_a$-functions and what their locations $\widetilde{\theta}_{a,i}$ at the end of the continuation are.\par
The key difficulty in those calculations is to figure out which path the TBA parameters $|m|$ and $C$ have to follow for a given path in terms of the cross ratios.
Basically, it amounts to solving the relations (\ref{eq:crs_yfcts}) numerically along every step of the continuation, with technical details described in \cite{Bartels:2014mka}.
Along every step of the continuation we then solve the equations $\widetilde{\Yf}_a(\theta)=-1$ numerically to see whether any of those solutions cross the real axis.
If this is the case, we rewrite the TBA equations as indicated in eq.~(\ref{eq:mod_TBA}) before we proceed with the analytic continuation.\par
While this algorithm involves a numerical analysis, this does not mean that our results are bound to any numerical accuracy.
The reason for this is that we can determine the endpoints of the solutions that have crossed the real axis exactly: at the endpoint of the continuation we go to the multi-Regge regime $|m|'\rightarrow\infty$, where we can neglect the integrals in the TBA equations, as explained in section \ref{sec:calc_gen}.
Therefore, at the endpoint of the continuation, the $\widetilde{\Yf}_a$-functions can be evaluated at the locations of the crossed solutions which by definition yields
\begin{equation}
	-1=\widetilde{\Yf}_a(\widetilde{\theta}_{a,i})=e^{-m_a'\cosh(\widetilde{\theta}_{a,i})-C_a'}\prod\limits_{a'}\prod\limits_{j=1}^{n_{a'}}S_{aa'}(\widetilde{\theta}_{a,i}-\widetilde{\theta}_{a',\,j})^{-\mathrm{sign}(\mathrm{Im}(\widetilde{\theta}_{a',\,j}))}.
	\label{eq:BA_eqs}
\end{equation}
This is a set of standard Bethe ansatz equations which can be solved exactly for the locations $\widetilde{\theta}_{a,i}$.
Therefore, our final result for the remainder function will be exact, even though it involves numerical intermediate steps.\par
\begin{figure}
	\centering
	\includegraphics[scale=1.3]{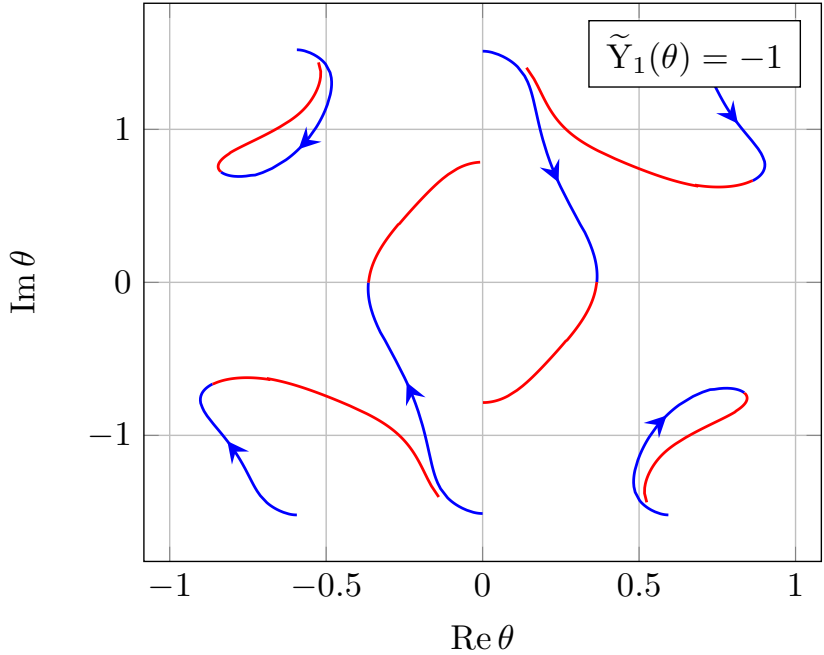}
	\caption{Movement of the solutions $\widetilde{\Yf}_1(\theta)=-1$ during the analytic continuation (\ref{eq:def_regge_region}). The values for the TBA parameters at the starting point of the continuation are chosen to be $|m|=10$, $w=1$ and $C=i\pi+\frac{3}{5}$. We switch colors from blue to red when two of the solutions cross the real axis. The convergence of the endpoint position of the crossing solutions against $\pm i\frac{\pi}{4}$ is clearly visible.}
	\label{fig:crossing_y1}
\end{figure}

Following this algorithm for the path (\ref{eq:def_regge_region}) for the cross ratios, we find that two solutions of the equation $\widetilde{\Yf}_1(\theta)=-1$ cross the real axis\footnote{The fact that we find crossing solutions for $\widetilde{\Yf}_1$ is related to our choice of $\mathrm{Re}\,(\widetilde{C})>0$ in our numerical analysis. Choosing $\mathrm{Re}\,(\widetilde{C})<0$ would lead to crossing solutions in $\widetilde{\Yf}_3$, which, however, gives rise to the same result for the remainder function.}, as shown in figure \ref{fig:crossing_y1}.
No solutions of the other $\widetilde{\Yf}_a$-functions cross the real axis.
We therefore have two crossing solutions, which we will call $\theta_\pm$ in the following.
Furthermore, solving the Bethe ansatz equations (\ref{eq:BA_eqs}) we find that 
\begin{equation}
	\theta_\pm=\pm i\frac{\pi}{4},
	\label{eq:endpoints_tpm}
\end{equation}
which confirms the numerical analysis shown in figure \ref{fig:crossing_y1}.
Note that the crossing picture is exactly the same as in the $\ttf$ -- case \cite{Bartels:2013dja}.
We have now assembled all necessary ingredients to calculate the remainder function in the Mandelstam region.

\subsection{$R_{\ttt}$ in the Mandelstam region and subleading kinematics}
\label{sec:subleading_TBA}
Having discussed the analytic continuation of the TBA and the resulting crossing solutions, we now proceed to calculate the remainder function in the Mandelstam region.
Our treatment will be similar to that in \cite{Bartels:2010ej, Bartels:2013dja}, but more careful, as the key point of this paper is the determination of contributions which are kinematically suppressed and which were not considered in those references.\par
We begin from the modified TBA equations valid in the Mandelstam region.
Since we send $|m|'\rightarrow\infty$ at the end of the continuation, we can neglect all integrals and find
\begin{equation}
	\log\widetilde{\Yf}'_a(\theta)=-m'_a\cosh\theta-C'_a+\log\left(\frac{S_{a1}(\theta+i\frac{\pi}{4})}{S_{a1}(\theta-i\frac{\pi}{4})}\right).
	\label{eq:mod_TBA_3t3}
\end{equation}
From the $\widetilde{\Yf}'_a$-functions we can then calculate the cross ratios at the endpoint of the continuation through the relations (\ref{eq:crs_yfcts}), from which we find
\begin{equation}
	u_2'=\left(1-\frac{2\sqrt{2}}{\sqrt{2}-\cos\varphi'+\sin\varphi'}\right)\varepsilon' w',\quad
	u_3'=\left(1+\frac{2\sqrt{2}}{-\sqrt{2}+\cos\varphi'+\sin\varphi'}\right)\frac{\varepsilon'}{w'},
	\label{eq:crs_ac}
\end{equation}
where we defined $\varepsilon':=e^{-|m|'\cos\varphi'}$ and $w':=e^{|m|'\sin\varphi'}$.
Eqs.~(\ref{eq:crs_ac}) are valid up to corrections of $\mathcal{O}({\varepsilon'}^{\,2})$.
Using our choice of path (\ref{eq:def_regge_region}) we then demand that 
\begin{equation}
	u_2'=-u_2,\quad u_3'=-u_3,
	\label{eq:ident_crs}
\end{equation}
which we can solve to obtain relations between the new parameters $\varepsilon'$, $w'$ and the parameters $\varepsilon$ and $w$.
Using $\varphi'=\tan^{-1}\left(-\frac{\log w'}{\log\varepsilon'}\right)$, these equations can be solved order by order in $\frac{1}{\log\varepsilon}$ and we find
\begin{align}
	\varepsilon'=&-\frac{1}{\gamma}\varepsilon\left(1-\sqrt{2}\,\frac{\log^2w}{\log^2\varepsilon}+\mathcal{O}(\log^{-3}\varepsilon)\right),	\label{eq:TBA_params_ac}\\
	w'=&\,w\left(1-2\sqrt{2}\,\frac{\log w}{\log\varepsilon}+4\frac{\log^2w}{\log^2\varepsilon}+4\sqrt{2}(\sqrt{2}-\log(1+\sqrt{2}\,))\frac{\log w}{\log^2\varepsilon}+\mathcal{O}(\log^{-3}\varepsilon)\right),\nonumber
\end{align}
where $\gamma=-3-2\sqrt{2}$.
These are the subleading kinematic corrections we are after in this paper.
In the previous analysis, only the leading terms (i.e. without any factors of $\log^{-n}\varepsilon$) were considered.
It is easy to see that these are the dominant corrections in the limit $\varepsilon\rightarrow 0$, since both the integrals we neglect in the TBA equations as well as the higher order terms neglected in eq.~(\ref{eq:crs_ac}) are of the form $\mathcal{O}(\varepsilon^{n})$ and therefore much smaller than the corrections considered here\footnote{Note that for any given numerical value of $\varepsilon$ there is, of course, an exponent $N$ such that $\log^{-N}\varepsilon$ and $\varepsilon$ are of the same order. However, here we are interested in the formal limit where $\varepsilon$ is arbitrarily small, when indeed all corrections of the form $\log^{-n}\varepsilon$ are smaller than the corrections of the form $\varepsilon^n$.}.
Note that the relations (\ref{eq:TBA_params_ac}), together with the exact relation for $C$ (\ref{eq:c_crs}), are also compatible with the third condition on the cross ratios, $u_1'=u_1$. 
Let us now examine how those subleading terms affect the remainder function.\par
%%%%%%%%%%%
The $\Delta$ -- contribution to the remainder function is easily evaluated, since it is a function of the cross ratios and we can immediately determine the behavior during the continuation (\ref{eq:def_regge_region}) to find
\begin{equation}
	\Delta'=\Delta+\frac{\pi^2}{4}+i\frac{\pi}{2}\log\left[-\left(w+\frac{1}{w}+2\cosh C\right)\right].
	\label{eq:delta_ac}
\end{equation}
Note that due to the behavior (\ref{eq:mrl_crs}) and the parametrization (\ref{eq:param_tba}) the argument of the logarithm in eq.~(\ref{eq:delta_ac}) is positive.
For $A_{\mathrm{per}}$, we use eq.~(\ref{eq:TBA_params_ac}) and find\footnote{Note that due to the quadratic term $\log^2\varepsilon'$ in $A_{\mathrm{per}}$ we always have to expand the parameters $\varepsilon'$, $w'$ in eq.~(\ref{eq:TBA_params_ac}) one order higher than the order we want to compute the remainder function to.}
\begin{equation}
	A_{\mathrm{per}}'=\frac{1}{4}\log^2\varepsilon'+\frac{1}{4}\log^2 w' =A_{\mathrm{per}}-\log(1+\sqrt{2}\,)+\log^2(1+\sqrt{2}\,)-\frac{3}{\sqrt{2}}\frac{\log^2 w}{\log\varepsilon}.
	\label{eq:aper_ac}
\end{equation}
Lastly, the $A'_{\mathrm{free}}$-contribution can be calculated via eq.~(\ref{eq:mod_a_free_per}) once we specify the number and endpoints of the crossing solutions.
After neglecting the integrals, we obtain
\begin{equation}
	A_{\mathrm{free}}'=-\sqrt{2}\,|m|'=\sqrt{2}\log\varepsilon'\sqrt{1+\frac{\log^2 w'}{\log^2\varepsilon'}}=\sqrt{2}\,\log\varepsilon-2\sqrt{2}\,\log(1+\sqrt{2}\,)+\frac{1}{\sqrt{2}}\frac{\log^2 w}{\log\varepsilon}.
	\label{eq:afree_ac}
\end{equation}
Note again that as described in section \ref{sec:calc_gen} the integrals in $A'_{\mathrm{free}}$ are of $\mathcal{O}(\varepsilon \log\varepsilon)$ and therefore much smaller than the corrections determined above.
We can now simply add up all contribution to find the remainder function
\begin{align}
	R_{\ttt}-i\pi\delta=&-\frac{\sqrt{\lambda}}{2\pi}\left[\left(\sqrt{2}-\log(1+\sqrt{2}\,)\right)\log\varepsilon+\frac{\pi^2}{4}-2\sqrt{2}\,\log(1+\sqrt{2}\,)+\log^2(1+\sqrt{2}\,)\right.\nonumber\\
	&\left.\quad-\sqrt{2}\,\frac{\log^2 w}{\log\varepsilon}+\mathcal{O}(\log^{-2}\varepsilon)\right],
	\label{eq:r6_ac}
\end{align}
where
\begin{equation}
	\delta=-\frac{\sqrt{\lambda}}{4\pi}\log\left[-\left(w+\frac{1}{w}+2\cosh C\right)\right]=\frac{\sqrt{\lambda}}{8\pi}\log\tilde{u}_2\tilde{u}_3.
	\label{eq:ttt_phase}
\end{equation}
This procedure can, of course, easily be automatized and we have obtained the subleading corrections to ten orders.
While the explicit form of subleading coefficients is not particularly illuminating, we can now examine the BFKL OPE of \cite{Basso:2014pla} and try to obtain the same coefficients from this approach.

%%%%%%%%%%%%%%%%%%%%%%%%%%%%%%%%%%%%%%%%%%%%%%%%%%%%%%%%%%%%%%%%%%%%%%%%%%%%%%%

\section{Subleading kinematics from the OPE}
\label{sec:subleading_ope}
As mentioned in the introduction, a conjecture for a finite-coupling expression for the six-point remainder function in the multi-Regge limit is put forward in \cite{Basso:2014pla}.
It is given by
\begin{equation}
	e^{R_{\ttt}-i\pi\delta}=-2\pi i \sum\limits_{m=-\infty}^{\infty}(-1)^me^{im\phi}\int\limits_{-\infty}^{\infty}\frac{du}{2\pi}\mu^{\mathrm{BFKL}}_m(u) e^{i(\sigma-\tau)\nu(u,m)+(\sigma+\tau)\,\omega(u,m)},
	\label{eq:bfkl_ope}
\end{equation}
where the kinematic variables are related to the cross ratios via
\begin{equation}
	\tau+\sigma=-\frac{1}{2}\log u_2u_3,\,\,\sigma-\tau=\frac{1}{2}\log \frac{u_2}{u_3}.
	\label{eq:def_tau_sigma}
\end{equation}
Having in mind a comparison with the TBA result, we can use eq.~(\ref{eq:param_tba}) to trade the variables $\sigma$, $\tau$ for the TBA variables and find
\begin{equation}
	\tau+\sigma=-\log\varepsilon,\,\,\sigma-\tau=\log w.
	\label{eq:rel_params_tba_ope}
\end{equation}
Furthermore, $\delta$ in eq.~(\ref{eq:bfkl_ope}) agrees with the phase of the TBA result (\ref{eq:ttt_phase}).
The finite coupling coupling conjectures for the BFKL eigenvalue $\omega(m, u)$, $\nu(m,u)$ and the impact factor (or BFKL measure, both expressions will be used interchangeably) $\mu^{\mathrm{BFKL}}_m(u)$ are given in \cite{Basso:2014pla}.
Here, we only spell out their form in the strong coupling limit $\sqrt{\lambda}\rightarrow\infty$.
The strong coupling expressions for the BFKL eigenvalue $\omega(m,u)$ and $\nu(m,u)$ are already derived in \cite{Basso:2014pla} and read
\begin{equation}
\begin{aligned}
	\omega(\theta)\big|_{\mathrm{SC}} &= \frac{\sqrt{\lambda}}{4\pi}\left[\frac{2\sqrt{2}\cosh\theta}{\cosh 2\theta}-\log\left(\frac{\sqrt{2}\cosh\theta+1}{\sqrt{2}\cosh\theta-1}\right)\right],\\
	\nu(\theta)\big|_{\mathrm{SC}} &= \frac{\sqrt{\lambda}}{4\pi}\left[\frac{2\sqrt{2}\sinh\theta}{\cosh 2\theta}-i\log\left(\frac{1+i\sqrt{2}\sinh\theta}{1-i\sqrt{2}\sinh\theta}\right)\right],
\end{aligned}
	\label{eq:bfkl_sc}
\end{equation}
where $\uh:=u/(2g)=:\tanh(2\theta)$, with $g=\sqrt{\lambda}/(4\pi)$.
Eqs.~(\ref{eq:bfkl_sc}) hold for $\uh<1$.
Similar expressions hold for $\uh>1$ but will not be needed in the following, they can be found in \cite{Basso:2014pla}.
Note that there is no $m$-dependence in eq.~(\ref{eq:bfkl_sc}), which is a feature of the leading order at strong coupling.
To extract the subleading kinematical corrections, however, we also need the behavior of the impact factor $\mu^{\mathrm{BFKL}}(u)$ at strong coupling.
This quantity is not derived in \cite{Basso:2014pla}.
We fill this gap by using the known strong coupling result of a formally related object, the OPE measure of the small fermion\footnote{We would like to point out that the result (\ref{eq:fin_res_measure_mt}), while unpublished, was already derived by Benjamin Basso and thank him for sharing the final expression (\ref{eq:fin_res_measure_mt}) with us.}.
Given the technical nature of the derivation, we provide the details in appendix \ref{sec:impact_factor} and simply quote the result here, which reads
\begin{align}
	\log\mu^{\mathrm{BFKL}}(\theta)\big|_{\mathrm{SC}}=&\frac{\sqrt{\lambda}}{2\pi}\Bigg[\int\limits_{\theta}^\infty\int\limits_\theta^\infty \frac{d\theta_1 d\theta_2}{\cosh(2\theta_1)\cosh(2\theta_2)}\frac{2}{\cosh(\theta_1-\theta_2)}+\frac{\pi}{2}\left(\frac{\nu(\uh)}{g}-2\uh\right)+\pi-\frac{\pi^2}{2}\Bigg],
	\label{eq:fin_res_measure_mt}
\end{align}
which holds in the region $\uh<1$.
\subsection{Extracting the subleading pieces}
Since all quantities in eqs.~(\ref{eq:bfkl_sc}, \ref{eq:fin_res_measure_mt}) scale like $\sqrt{\lambda}$ we can evaluate the integral by means of a saddle point approximation, i.e.\ we first solve the equation
\begin{equation}
	0=\partial_\theta\log\mu^{\mathrm{BFKL}}(\theta)+i\log w\,\partial_\theta\,\nu(\theta)-\log\varepsilon\,\partial_\theta\,\omega(\theta)
	\label{eq:saddle_point}
\end{equation}
for the saddle point $\theta_0$ and then obtain the remainder function as
\begin{equation}
	e^{R_{\ttt}-i\pi\delta}\sim e^{\log\mu^{\mathrm{BFKL}}(\theta_0)+i\log w\,\nu(\theta_0)-\log\varepsilon\,\omega(\theta_0)}.
	\label{eq:rem_saddle_point}
\end{equation}
Recall that we have lost the $m$-dependence in going to the strong coupling limit and do not know which $m$-mode is dominant.
We therefore write $\sim$ in eq.~(\ref{eq:rem_saddle_point}).
Similar to the TBA-case, we can expand the result in orders of $\frac{1}{\log\varepsilon}$.
For example, it is easy to see that the leading order result is given by
\begin{equation}
	\theta_0=0+\mathcal{O}(\log^{-1}\varepsilon)
	\label{}
\end{equation}
which leads to
\begin{equation}
	R_{\ttt}-i\pi\delta = -\frac{\sqrt{\lambda}}{2\pi}\left[\log\varepsilon \left(\sqrt{2}-\log(1+\sqrt{2}\,)\right)+\frac{\pi^2}{4}-2\sqrt{2}\log(1+\sqrt{2}\,)+\log^2(1+\sqrt{2}\,)) \right],
	\label{}
\end{equation}
which precisely agrees with the leading order result of the TBA calculation, see eq.~(\ref{eq:r6_ac}).
We have performed this calculation up to ten orders in $\log^{-n}\varepsilon$ and found perfect matching of the BFKL OPE and the TBA results, which strongly supports the conjectured finite-coupling expressions of the BFKL OPE.
Given this result, it is natural to ask whether the two formalisms can be directly mapped onto each other at strong coupling.
We will show in the next section that this is indeed the case.

%%%%%%%%%%%%%%%%%%%%%%%%%%%%%%%%%%%%%%%%%%%%%%%%%%%%%%%%%%%%%%%%%%%%%%%%%%%%%%%

\subsection{Mapping TBA $\leftrightarrow$ BFKL OPE}
\label{sec:matching}
For both the TBA and the BFKL OPE there are two steps involved in determining the remainder function -- first the kinematic aspect of finding the saddle point for the OPE and finding the parameters at the endpoint $\varepsilon'$ and $w'$ for the TBA and then evaluating the remainder function on these solutions.

It is therefore natural to expect that the saddle point equation (\ref{eq:saddle_point}) can be mapped to the equation determining $\varphi'$ (\ref{eq:ident_crs}).
To see this, we start from the definition of $\varphi'$,
\begin{equation}
	\varphi'=\tan^{-1}\left(-\frac{\log\varepsilon'}{\log w'}\right)
\end{equation}
and use eqs.~(\ref{eq:crs_ac}, \ref{eq:ident_crs}, \ref{eq:param_tba}) to rewrite this in the form
\begin{equation}
	i\log w\,g_1(\varphi')-\log\varepsilon\,g_2(\varphi')+g_3(\varphi')=0,
	\label{eq:struc_sp_tba}
\end{equation}
where $g_i(\varphi')$ are some functions, whose exact form is not illuminating.
This already has the same structure as the saddle point equation for $\theta_0$, see eq.~(\ref{eq:saddle_point}).
Of course, we can always multiply eq.~(\ref{eq:struc_sp_tba}) by an overall factor.
Fixing this factor by comparing the coefficients of $\log w$ for eq.~(\ref{eq:struc_sp_tba}) and eq.~(\ref{eq:saddle_point}), we find that also the other coefficient functions match perfectly, once we identify
\begin{equation}
	\theta_0=i \varphi'.
	\label{eq:ident_param_tba_ope}
\end{equation}\par
%%%%%%%%%%%%%%%%%%%%%%%%%%%
Similarly, it should then be possible to map the expressions for the remainder function (\ref{eq:rem_saddle_point}) and (\ref{eq:def_r_sc}) onto each other.
We begin by noting that the remainder function on the TBA side can be written as
\begin{equation}
	e^{R_{\ttt}-i\pi\delta}\sim e^{-\frac{\sqrt{\lambda}}{2\pi}\left(-\sqrt{2}|m|'+\frac{1}{4}{|m|'}^{\,2}-\frac{1}{4}|m|^2+\frac{\pi^2}{4}\right)}
	\label{eq:rem_tba}
\end{equation}
in terms of the parameters $|m|'$ and $|m|$.
Keeping in mind the identification (\ref{eq:ident_param_tba_ope}) as well as the structure of the remainder function on the OPE side (\ref{eq:rem_saddle_point}), we use eqs.~(\ref{eq:ident_crs}, \ref{eq:param_tba}) to rewrite the TBA remainder function (\ref{eq:rem_tba}) in the form
\begin{equation}
	e^{R_{\ttt}-i\pi\delta}\sim e^{\frac{\sqrt{\lambda}}{2\pi}\left(i\log w\,h_1(\varphi')-\log\varepsilon\,h_2(\varphi')+h_3(\varphi')\right)}.
	\label{eq:rem_tba_rewritten}
\end{equation}
This does not immediately reproduce eq.~(\ref{eq:rem_saddle_point}) due to a small subtlety -- we have some freedom in rearranging terms due to the relation
\begin{equation}
	\log w = -\tan\varphi' \log\varepsilon+h_5(\varphi'),
	\label{eq:res_freedom}
\end{equation}
which can be derived from eqs.~(\ref{eq:crs_ac}, \ref{eq:ident_crs}) and where $h_5(\varphi')$ is a complicated expression, which we do not spell out explicitly.
We then obtain
\begin{equation}
	e^{R_{\ttt}-i\pi\delta}\sim e^{\frac{\sqrt{\lambda}}{2\pi}\left(i\log w (h_1(\varphi')+h_4(\varphi'))-\log\varepsilon (h_2(\varphi')-i h_4(\varphi')\tan\varphi')+(h_3(\varphi')-ih_4(\varphi')h_5(\varphi'))\right)},
	\label{eq:rem_TBA_final}
\end{equation}
where $h_4(\varphi')$ is an arbitrary function of $\varphi'$.
As for the saddle point equation, we match the coefficient of $\log w$ with the OPE result (\ref{eq:rem_saddle_point}) to fix $h_4(\varphi')$ in eq.~(\ref{eq:rem_TBA_final}).
We then again find perfect agreement for the other two coefficient functions.
This nicely shows that the two formalisms are identical at strong coupling.

%%%%%%%%%%%%%%%%%%%%%%%%%%%%%%%%%%%%%%%%%%%%%%%%%%%%%%%%%%%%%%%%%%%%%%%%%%%%%%%

\section{Subleading kinematics for the $2\rightarrow 5$ -- amplitude}
\label{sec:ttf}
As another application of the subleading kinematic corrections let us consider the seven-point remainder function.
In this case, we have six independent cross ratio, which behave as
\begin{alignat}{3}
	& u_{11}=1-\varepsilon_2\left(w_2+\frac{1}{w_2}+2\cosh C_2\right), \quad && u_{21}=\varepsilon_2 w_2,\quad && u_{31}=\frac{\varepsilon_2}{w_2},\nonumber\\
	& u_{12}=1-\varepsilon_1\left(w_1+\frac{1}{w_1}+2\cosh C_1\right), \quad && u_{22}=\varepsilon_1 w_1,\quad && u_{32}=\frac{\varepsilon_1}{w_1}\label{eq:crs7pt}
\end{alignat}
in the multi-Regge limit where $\varepsilon_i\rightarrow 0$, the $w_i$ are real and constant, and $C_i$ are purely imaginary and constant.
Corrections to the cross ratios in eq.~(\ref{eq:crs7pt}) are of $\mathcal{O}(\varepsilon^2)$.
Furthermore, there is another, dependent cross ratios $\tilde{u}$, which behaves as $1-\tilde{u}\sim\varepsilon^2$ in the multi-Regge limit and which is connected with the independent cross ratios via a conformal Gram relation, see \cite{Bartels:2014mka} for details.
In the seven-point case, there are several interesting Mandelstam regions.
Here, we focus on the Mandelstam region, which is probed by the analytic continuation
\begin{equation}
	\tilde{u}\rightarrow e^{-2\pi i} \tilde{u},
\end{equation}
with all other cross ratios held fixed.
Subtleties in probing this region, usually denoted as $P_{7,---}$, from the TBA are discussed in \cite{Bartels:2014mka}, but do not play a role here.
In this region, the all-loop remainder function is expected to be of the form \cite{Bartels:2014jya}
\begin{align}
	&e^{R_{7,---}+i\delta_{7,---}}=\,i\lambda\sum\limits_{n_1,n_2}(-1)^{n_1+n_2}\left(\frac{z_1}{z_1^\ast}\right)^{\frac{n_1}{2}}\left(\frac{z_2}{z_2^\ast}\right)^{\frac{n_2}{2}}\int\frac{d\nu_1 d\nu_2}{(2\pi)^2}\,\varPhi(\nu_1,n_1)^\ast |z_1|^{2i\nu_1}\label{eq:7pt_disprel}\\
	&\quad\times\big(-\sqrt{u_{21}u_{31}}\,\big)^{-\omega(\nu_1,n_1)}C(\nu_1,n_1,\nu_2,n_2)\big(-\sqrt{u_{22}u_{32}}\,\big)^{-\omega(\nu_2,n_2)}|z_2|^{2i\nu_2}\varPhi(\nu_2,n_2)\big|_{\mathrm{sub}}+\dots,\nonumber
\end{align}
where the subscript \textit{sub} means that the one-loop contribution has been subtracted and the dots indicate phases and Regge pole contributions which play no role in following strong coupling discussion.
The relation between the $z_i$ in eq.~(\ref{eq:7pt_disprel}) and our parameters is given by $z_i=\frac{1}{w_{3-i}}e^{C_{3-i}}$.
While the BFKL eigenvalue $\omega(\nu,n)$ and the impact factors $\varPhi(\nu,n)$ in eq.~(\ref{eq:7pt_disprel}) are the same as in the six-point case, the central emission vertex $C(\nu_1,n_1,\nu_2,n_2)$ is a new ingredient in the seven-point case, which links the two integrations.
The form of the remainder function in eq.~(\ref{eq:7pt_disprel}) is supported by the explicit result in LLA \cite{Bartels:2011ge} and although higher logarithmic orders are still unknown, it is conceivable that those just introduce corrections to the BFKL eigenvalue and the impact factor, as well as the central emission vertex.
In fact, the remainder function in this Mandelstam region at strong coupling is investigated in \cite{Bartels:2014mka} with the result that
\begin{equation}
	e^{R_{7,---}+i\tilde{\delta}_{7,---}}\sim\left(\sqrt{u_{21}\,u_{31}\,u_{22}\,u_{32}}\right)^{\frac{\sqrt{\lambda}}{2\pi}e_2},\label{eq:7pt_sc_res}
\end{equation}
where again $e_2=-\sqrt{2}+\log\left(1+\sqrt{2}\right)$ and
\begin{equation}
	\tilde{\delta}_{7,---}=\frac{\sqrt{\lambda}}{4}\log\left(\frac{\sqrt{u_{21}\,u_{31}\,u_{22}\,u_{32}}}{1-\tilde{u}}\right)+\frac{\sqrt{\lambda}}{4}\log\left(\frac{u_{21}u_{32}}{u_{31}u_{22}}\right)=\delta_{7,---}+\frac{\sqrt{\lambda}}{4}\log\left(\frac{u_{21}u_{32}}{u_{31}u_{22}}\right).\label{eq:7pt_phase}
\end{equation}
Note that the phase $\tilde{\delta}_{7,---}$ in eq.~(\ref{eq:7pt_phase}) already slightly differs from the predicted valued $\delta_{7,---}$ in \cite{Bartels:2013jna, Bartels:2014jya} by an additional piece.
This difference, however, could well arise from the contribution of the central emission vertex at strong coupling.
The result (\ref{eq:7pt_sc_res}) is clearly compatible with the form (\ref{eq:7pt_disprel}).
Assuming that the form (\ref{eq:7pt_disprel}) holds at strong coupling, we can make some statements regarding the central emission vertex at strong coupling. 
The result (\ref{eq:7pt_sc_res}) clearly factorizes in the two triplets of cross ratios, which seems to suggest that the central emission vertex is trivial (up to a potential phase which depends on the cross ratios, as mentioned above) at strong coupling.
However, following the strategy outlined in the previous sections for the six-point case, we can analyze the subleading kinematic contributions and see if they factorize in the two triplets, as well, or if there are terms that couple the triplets.
In the latter case, this would mean that the central emission vertex is not trivial at strong coupling and links the two integrations.\par
While the individual contributions to the remainder function from the TBA perspective are different from the six-point case, the calculation of the subleading terms proceeds in exactly the same way.
We therefore refrain from going through the calculation and directly present our result.
All formulas necessary for the derivation are presented in \cite{Bartels:2014mka}.
We parametrize the remainder function as
\begin{equation}
	R_{7,---}+i\tilde{\delta}_{7,---}=\sum\limits_{k_1=-1}^{\infty}\sum\limits_{k_2=-1}^{\infty} c_{k_1,k_2}(w_1,w_2)\log^{-k_1}\varepsilon_1\log^{-k_2}\varepsilon_2.
\end{equation}
In this notation, the leading terms of eq.~(\ref{eq:7pt_sc_res}) correspond to the terms $c_{-1,0}$, $c_{0,-1}$ and $c_{0,0}$.
Some of the lowest subleading terms read\footnote{We provide the full form of the first four orders of subleading terms in the file \texttt{7pt\_subleading.m} attached to the arXiv submission of this paper.}
\begin{align}
	c_{1,0}(w_1,w_2)=&-\sqrt{2} \log ^2w_1-\left(2\sqrt{2}\log \left(1+\sqrt{2}\right)-2+3 i \sqrt{2} \pi\right)\log w_1+\mathrm{const.}\,,\\
	c_{0,1}(w_1,w_2)=&\,\,c_{1,0}\left(1/w_2,1/w_1\right)\quad\mathrm{via\,\,target\text{-}projectile\,\,symmetry\,\,and}\\
	c_{1,1}(w_1,w_2) =& -6 \log w_1 \log w_2+\left(6 \log \left(1+\sqrt{2}\right)+9i\pi-2\sqrt{2}\right)\log w_1\nonumber\\&\quad-\left(6 \log \left(1+\sqrt{2}\right)+9i\pi-2\sqrt{2}\right)\log w_2+\mathrm{const.}
\end{align}
While their explicit form is not particularly simple, it is very interesting that we find subleading terms of the form $\log^{-1}\varepsilon_1\,\log^{-1}\varepsilon_2$.
These terms couple the two triplets of cross ratios which indicates that the integrals in eq.~(\ref{eq:7pt_disprel}) are still coupled at strong coupling.
This, in turn, implies that the central emission vertex does not become trivial at strong coupling, assuming that the form (\ref{eq:7pt_disprel}) holds.
\section{Conclusions}
\label{sec:conclusions}
In this paper, we have re-examined the six-point remainder function in the multi-Regge limit at strong coupling and found that there is an infinite set of kinematically subleading corrections which were not considered in previous publications and which allow a detailed comparison with the finite-coupling expressions derived from the Wilson loop OPE.
After comparing these subleading pieces order by order, we have shown that the two frameworks can actually be precisely mapped onto each other at strong coupling.\par
We then studied the corresponding calculation for the seven-point remainder function in the Mandelstam region $P_{7,---}$ and found that there are subleading terms which couple the two triplets of cross ratios.
We interpret these terms as coming from the contribution of the central emission vertex which therefore cannot be trivial at strong coupling.
Once a finite coupling prediction for the central emission vertex becomes available, it would be interesting to check it against our TBA calculation.
It would be also interesting to see if it is possible to constrain the form of the strong coupling limit of the central emission vertex using our results.
Furthermore, it would be interesting to see what happens beyond the leading order at strong coupling, when the degeneracy of the different $m$-modes of eq.~(\ref{eq:bfkl_ope}) is lifted, from which we could get a more precise picture of which mode is dominant at strong coupling.
This is more difficult than the case considered in this paper and is left as an open question for future investigations.

\acknowledgments
I would like to thank Benjamin Basso, Simon Caron-Huot, Ben Hoare and Amit Sever for many valuable discussions as well as Benjamin Basso, Elli Pomoni and Maikel de Vries for helpful comments on the draft.
My work is partially supported by the Swiss National Science Foundation through the NCCR SwissMAP.
\appendix

\section{Derivation of the impact factor at strong coupling}
\label{sec:impact_factor}
In this appendix, we present the technical details on the derivation of the impact factor at strong coupling, eq.~(\ref{eq:fin_res_measure_mt}).
We are interested in the limit $g\rightarrow\infty$, while keeping the rescaled rapidity $\uh=\frac{u}{2g}$ fixed and the mode number $m$ of $\mathcal{O}(1)$.
To derive the impact factor at strong coupling, we start from the finite-coupling expression for the impact factor as derived in \cite{Basso:2014pla},
\begin{equation}
	\mu_m^\mathrm{BFKL}(u)=\frac{g^2(x^{[+m]}x^{[-m]}-g^2)}{x^{[+m]}x^{[-m]}\sqrt{\left(x^{[+m]}x^{[+m]}-g^2\right)\left(x^{[-m]}x^{[-m]}-g^2\right)}}e^{A+2f^{(3)}_{\mathrm{BFKL},m}(u)-2f^{(4)}_{\mathrm{BFKL},m}(u)},
	\label{eq:fc_measure}
\end{equation}
where
\begin{equation}
	\begin{aligned}
	x^{[\pm m]}&=x\left(u\pm i\frac{m}{2}\right),\quad\mathrm{with}\\
	x(u)&=\frac{1}{2}(u+\sqrt{u^2-4g^2}).
	\end{aligned}
	\label{eq:def_xu}
\end{equation}
The constant $A$ is given by
\begin{equation}
	A=2\int\limits_0^\infty\frac{dt}{t}\frac{1-J_0(2gt)^2}{e^t-1}-\frac{\pi^2}{4}\Gamma_{\mathrm{cusp}},
	\label{eq:def_A}
\end{equation}
while the functions $f^{(3)}_{\mathrm{BFKL},m}(u)$, $f^{(4)}_{\mathrm{BFKL},m}(u)$ are defined via the infinite-dimensional matrices
\begin{equation}
	\mathbb{K}_{ij}=2j(-1)^{j(i+1)}\int\limits_0^\infty\frac{dt}{t}\frac{J_i(2gt)J_j(2gt)}{e^t-1},\quad \mathbb{M}=(\mathbbm{1}+\mathbb{K})^{-1},\quad \mathbb{Q}_{ij}=\delta_{ij}(-1)^{i+1}i
	\label{eq:def_mat}
\end{equation}
and the source terms
\begin{equation}
\begin{aligned}
	\kappa^{\mathrm{BFKL}}_{m,j}&=-\int\limits_{0}^\infty\frac{dt}{t}\frac{J_j(2gt)}{e^t-1}\left(\frac{e^{t\delta_j^{\mathrm{even}}}-(-1)^je^{t\delta_j^{\mathrm{odd}}}}{2}\cos(ut)e^{-mt/2}-J_0(2gt)\right),\\
	\widetilde{\kappa}^{\mathrm{BFKL}}_{m,j}&=-\int\limits_0^{\infty}\frac{dt}{t}\frac{J_j(2gt)}{e^t-1}\frac{e^{t\delta_j^{\mathrm{even}}}+(-1)^j e^{t\delta_j^\mathrm{odd}}}{2}\sin(ut)e^{-mt/2},
\end{aligned}
	\label{eq:source_bfkl}
\end{equation}
where $\delta^{\mathrm{even/odd}}_j:=\frac{1}{2}\left(1\pm(-1)^j\right)$.
In terms of these objects, the functions $f^{(3)}_{\mathrm{BFKL},m}(u)$, $f^{(4)}_{\mathrm{BFKL},m}(u)$ are defined as
\begin{equation}
\begin{aligned}
	f^{(3)}_{\mathrm{BFKL},m}(u)&=2\widetilde{\kappa}^{\mathrm{BFKL}}_m(u)\cdot\mathbb{Q}\cdot\mathbb{M}\cdot\widetilde{\kappa}^{\mathrm{BFKL}}_m(u)\quad\mathrm{and}\\
	f^{(4)}_{\mathrm{BFKL},m}(u)&=2\kappa^{\mathrm{BFKL}}_m(u)\cdot\mathbb{Q}\cdot\mathbb{M}\cdot\kappa^{\mathrm{BFKL}}_m(u).
\end{aligned}
	\label{eq:def_f3f4}
\end{equation}
The strong coupling limit $g\rightarrow\infty$ can be readily performed for all parts of eq.~(\ref{eq:fc_measure}) except for the functions $f^{(3)}_{\mathrm{BFKL},m}(u)$, $f^{(4)}_{\mathrm{BFKL},m}(u)$, which are more involved.
Rescaling $u\rightarrow\hat{u}$ and expanding at strong coupling, we easily see that
\begin{equation}
	\left.x^{[\pm m]}(u)\right|_{\mathrm{SC}}=\left.x(\uh)\right|_{\mathrm{SC}},
\end{equation}
where the subscript $\mathrm{SC}$ stands for the leading order at strong coupling.
Therefore, we immediately find that the leading term of the prefactor eq.~(\ref{eq:fc_measure}) is of order $g^0$.
Since we are only interested in the leading exponential behavior at strong coupling, we can therefore drop the prefactor from now on.
For the constant $A$ we make the substitution $t\rightarrow \frac{t}{2g}$ and expand at strong coupling to obtain
\begin{equation}
	\left.A\right|_{\mathrm{SC}}=\frac{\sqrt{\lambda}}{2\pi}\left(\frac{8}{\pi}-\frac{\pi^2}{4}\right),
	\label{eq:A_sc}
\end{equation}
where we used that $\Gamma_{\mathrm{cusp}}=\frac{\sqrt{\lambda}}{2\pi}$ at strong coupling.
Let us now turn to the functions $f_{\mathrm{BFKL},m}^{(3)}(u)$, $f_{\mathrm{BFKL},m}^{(4)}(u)$.
The definition of these functions given in eq.~(\ref{eq:def_f3f4}) is well-suited for an analysis at weak coupling, when the matrices appearing in eq.~(\ref{eq:def_mat}) can be truncated to finite size, since $\mathbb{K}_{ij}\sim g^{i+j}$ at weak coupling.
At strong coupling, however, all matrix entries are of the same order $\mathbb{K}_{ij}\sim g$ and only become numerically smaller as $i,j$ grow.
Therefore, one would have to work with the full infinite-dimensional matrices, which is not feasible.
%%%%%%%%%%%%%%
Fortunately, to leading order at strong coupling, we can make use of a formal similarity of the BFKL source terms to those of the small fermion excitation of the GKP string \cite{Basso:2010in}.
Indeed, at strong coupling we have
\begin{equation}
	\left.\widetilde{\kappa}^{\mathrm{BFKL}}_{m,j}(\hat{u})\right|_{\mathrm{SC}}=-2g\int\limits_0^\infty\frac{dt}{t^2}J_j(t)\frac{\left(1+(-1)^j\right)}{2}\sin(\hat{u}t).
	\label{eq:sc_kappat}
\end{equation}
As we can see from eq.~(\ref{eq:sc_kappat}), the result is independent of $m$, which reflects the known universality of the leading order result at strong coupling, which we also observed for the BFKL eigenvalue, cf.\ eq.~(\ref{eq:bfkl_sc}).
We will therefore drop the index for the strong coupling expressions in the following.
Upon taking the derivative with respect to $\uh$, we therefore have
\begin{equation}
	\left.\partial_{\uh}\widetilde{\kappa}^{\mathrm{BFKL}}_{j}\right|_{\mathrm{SC}}=-4g\left.\kappa^{\mathrm{SF}}_{j}(\hat{u})\right|_{\mathrm{SC}},
	\label{eq:der_kt}
\end{equation}
where $\mathrm{SF}$ stands for small fermion, see appendix B of \cite{Basso:2010in}.
Similarly we have that
\begin{equation}
	\left.\partial_{\hat{u}}\kappa^{\mathrm{BFKL}}_{j}\right|_{\mathrm{SC}}=4g\left.\widetilde{\kappa}^{\mathrm{SF}}_{j}(\hat{u})\right|_{\mathrm{SC}}.
	\label{eq:der_k}
\end{equation}
We can therefore use the strong coupling expansion of the functions $f_{\mathrm{SF}}^{(3,4)}$ for the small fermion case which are derived in \cite{Belitsky:2015qla} to obtain the corresponding expressions for the BFKL case.
As everything that follows only concerns the leading order in strong coupling, we will drop the subscript SC from now on.
\par
%%%%%%%%%%%%%%
We start from a slight generalization of the functions $f^{(3,4)}$, namely
\begin{equation}
\begin{aligned}
	f^{(3)}_{\mathrm{BFKL}}(\hat{u},\hat{v})&:=2\widetilde{\kappa}^{\mathrm{BFKL}}(\hat{u})\cdot\mathbb{Q}\cdot\mathbb{M}\cdot\widetilde{\kappa}^{\mathrm{BFKL}}(\hat{v}),\\
	f^{(4)}_{\mathrm{BFKL}}(\hat{u},\hat{v})&:=2\kappa^{\mathrm{BFKL}}(\hat{u})\cdot\mathbb{Q}\cdot\mathbb{M}\cdot\kappa^{\mathrm{BFKL}}(\hat{v}),
\end{aligned}
	\label{eq:def_f_uv}
\end{equation}
and equivalently for the small fermion functions $f^{(3,4)}_{\mathrm{SF}}(\hat{u},\hat{v})$.
We will determine these functions and take the limit $\hat{v}\rightarrow \hat{u}$ in the end.
As in the case of the BFKL eigenvalue $\omega(\hat{u})$ and $\nu(\hat{u})$, the resulting expressions take a different form in the regions $\hat{u},\hat{v}\lessgtr 1$, which we both examine in the following.

\subsection{$\hat{u},\hat{v}>1$}

The region $\hat{u}, \hat{v}>1$ is the natural kinematic regime for the small fermion excitation.
We can therefore immediately use the results for the small fermion measure at strong coupling as derived in \cite{Belitsky:2015qla}.
However, we will present some intermediate results of the derivation as they will be needed in the next section.
The derivation starts from an integral representation of the function $f^{(3)}_{\mathrm{SF}}(\hat{u},\hat{v})$\footnote{For the calculations in the small fermion case, we use the notation of \cite{Belitsky:2015qla} for all quantities with the exception that we change the subscript to SF to make the distinction with the BFKL quantities clear.}
\begin{equation}
	f^{(3)}_{\mathrm{SF}}(\hat{u},\hat{v})=\frac{1}{2}\int\limits_0^\infty\frac{d\tau}{\tau}\sin(\hat{v}\tau)\widetilde{\gamma}^\mathrm{f}_{-,\hat{u}}(\tau),
	\label{eq:int_f3}
\end{equation}
where the function $\widetilde{\gamma}^\mathrm{f}_{-,\hat{u}}(\tau)$ is shown to be given by 
\begin{equation}
	\widetilde{\gamma}^\mathrm{f}_{-,\hat{u}}(\tau)=\frac{\tau}{4g}\left[-\frac{1}{4}\left(\frac{\uh-1}{\uh+1}\right)^{\frac{1}{4}}W^+(\tau,\uh)+\frac{1}{4}\left(\frac{\uh+1}{\uh-1}\right)^{\frac{1}{4}}W^+(\tau, -\uh)\right]
	\label{eq:lo_gamma_f3_l}
\end{equation}
to leading order at strong coupling.
In eq.~(\ref{eq:lo_gamma_f3_l}), the function $W^+(\tau, \uh)$ is defined as
\begin{equation}
	W^+(\tau,\uh):=\frac{\sqrt{2}}{\pi}\int\limits_{-1}^{1}dk\left(\frac{1+k}{1-k}\right)^{\frac{1}{4}}\cos(\tau k)\frac{\mathcal{P}}{k-\uh},
	\label{eq:def_wp}
\end{equation}
where $\mathcal{P}$ denotes the principal value.
To carry out the integrals, we use the relation
\begin{equation}
	\int\limits_{0}^\infty d\tau \sin(\vh \tau)\cos(k\tau)=\frac{1}{2}\left(\frac{\mathcal{P}}{k+\vh}-\frac{\mathcal{P}}{k-\vh}\right).
	\label{eq:fouriercos}
\end{equation}
Due to the range of $k$ in the integration in eq.~(\ref{eq:def_wp}) and the assumption $\hat{u}, \hat{v}>1$, all principal value integrals become standard integrals and after partial fractioning, as well as using the identity
\begin{equation}
	\int\limits_{-1}^{1}\frac{dk}{\pi}\left(\frac{1+k}{1-k}\right)^{\frac{1}{4}}\frac{1}{k-p}=-\sqrt{2}\left(\frac{p+1}{p-1}\right)^{\frac{1}{4}}+\sqrt{2}
\end{equation}
for $|p|>1$, see e.g.\ \cite{Belitsky:2015qla},
we obtain the result
\begin{equation}
\begin{aligned}
	f^{(3)}_{\mathrm{SF}}(\uh,\vh)=-\frac{1}{32g}\Bigg[&\frac{1}{\uh+\vh}\left(\sqrtuhmg\sqrtvhmg-\sqrtuhpg\sqrtvhpg\right)\\
	&-\frac{1}{\uh-\vh}\left(\sqrtuhmg\sqrtvhpg-\sqrtuhpg\sqrtvhmg\right)\Bigg].
\end{aligned}
	\label{eq:res_f3_sf}
\end{equation}
Looking at the relations (\ref{eq:der_kt}, \ref{eq:der_k}) we see that we still need to re-integrate in $\hat{u}$, $\hat{v}$ to obtain the desired expression for the BFKL case.
As boundary values we use the relation
\begin{equation}
	\kappa^{\mathrm{BFKL}}(\hat{u})\cdot\mathbb{Q}\cdot\mathbb{M}\cdot\kappa^{\mathrm{BFKL}}(1)=\frac{\sqrt{\lambda}}{2\pi}\left(-\frac{\pi}{4}+\frac{2}{\pi}\right),
	\label{eq:bc_kappa_const}
\end{equation}
which can be easily obtained from identities presented in appendix A of \cite{Basso:2014pla} (see, in particular, eqs.~(A.10, A.17) in that reference) and holds to leading order at strong coupling.
We then obtain\footnote{In the second step we assume that summation and integration commute, which is supported by numerical checks.}
\begin{align}
	f^{(4)}_{\mathrm{BFKL}}(\uh,\vh)&=2\left(\int\limits^{\uh}_1 d\xi_1\frac{d\kappa^{\mathrm{BFKL}}}{d\xi_1}+\kappa^{\mathrm{BFKL}}(1)\right)\cdot\mathbb{Q}\cdot\mathbb{M}\cdot\left(\int\limits^{\vh}_1 d\xi_2\frac{d\kappa^{\mathrm{BFKL}}}{d\xi_2}+\kappa^{\mathrm{BFKL}}(1)\right)\nonumber\\
	&=16g^2\int\limits^{\uh}_1d\xi_1\int\limits^{\vh}_1d\xi_2 \,f^{(3)}_{\mathrm{SF}}(\xi_1,\xi_2)+\frac{\sqrt{\lambda}}{2\pi}\left(-\frac{\pi}{2}+\frac{4}{\pi}\right).
	\label{eq:res_f4_sf}
\end{align}
Taking the limit $\vh\rightarrow\uh$ then gives the expression $f^{(4)}_{\mathrm{BFKL}}(\uh)$ needed for the BFKL measure.\par
%%%%%%%%%%%%%%%%%%%%
Similarly the function $f^{(4)}_{\mathrm{SF}}(\hat{u},\hat{v})$ has an integral representation as
\begin{equation}
	f^{(4)}_{\mathrm{SF}}(\hat{u},\hat{v})=-\frac{1}{2}\int\limits_0^\infty\frac{d\tau}{\tau}\cos(\vh \tau)\gamma^{\mathrm{f}}_{+,\uh}(\tau),
	\label{eq:def_int_f4}
\end{equation}
where the function $\gamma^{\mathrm{f}}_{+,\uh}(\tau)$ is given by
\begin{equation}
	\gamma^{\mathrm{f}}_{+,\uh}(\tau)=\frac{\tau}{4g}\left[\frac{1}{4}\sqrtuhmg W^{-}(\tau,\uh)+\frac{1}{4}\sqrtuhpg W^{-}(\tau,-\uh)\right]
	\label{eq:lo_gamma_f4}
\end{equation}
to leading order at strong coupling.
Furthermore, the function $W^{-}(\tau,\uh)$ is defined as
\begin{equation}
	W^{-}(\tau,\uh):=\frac{\sqrt{2}}{\pi}\int\limits_{-1}^{1}dk\left(\frac{1+k}{1-k}\right)^{\frac{1}{4}}\sin(\tau k)\frac{\mathcal{P}}{k-\uh}.
	\label{eq:def_wm}
\end{equation}
Performing the integrals as before one obtains the result
\begin{equation}
\begin{aligned}
	f^{(4)}_{\mathrm{SF}}(\uh,\vh)=-\frac{1}{32g}\Bigg[&\frac{1}{\uh+\vh}\left(\sqrtuhmg\sqrtvhmg-\sqrtuhpg\sqrtvhpg\right)\\
	&+\frac{1}{\uh-\vh}\left(\sqrtuhmg\sqrtvhpg-\sqrtuhpg\sqrtvhmg\right)\Bigg].
\end{aligned}
\label{eq:res_f4_sf}
\end{equation}
To integrate this expression to the BFKL case we use the boundary values
\begin{equation}
	\begin{aligned}
	&\widetilde{\kappa}^{\mathrm{BFKL}}(1)\cdot\mathbb{Q}\cdot\mathbb{M}\cdot\widetilde{\kappa}^{\mathrm{BFKL}}(1)=\frac{\sqrt{\lambda}}{2\pi}\left(-\frac{\pi}{4}+\frac{\pi^2}{16}\right),\\
	&\widetilde{\kappa}^{\mathrm{BFKL}}(1)\cdot\mathbb{Q}\cdot\mathbb{M}\cdot\widetilde{\kappa}^{\mathrm{BFKL}}(\uh)=\frac{\pi}{8}\left(\nu(\uh)-4g\uh\right),
	\end{aligned}
	\label{eq:rels_f4}
\end{equation}
which are obtained similarly to eq.~(\ref{eq:bc_kappa_const}).
In eq.~(\ref{eq:rels_f4}), $\nu(\theta)$ is the corresponding expression of eq.~(\ref{eq:bfkl_sc}) for the region $\uh>1$ and is given by
\begin{equation}
	\nu(\theta)=\frac{\sqrt{\lambda}}{2\pi}\left[\frac{\pi}{2}+\frac{1}{\sinh\theta}+\frac{i}{2}\log\left(\frac{\sinh\theta+i}{\sinh\theta-i}\right)\right],
	\label{eq:def_nu_g1}
\end{equation}
see \cite{Basso:2014pla}.
We then find
\begin{equation}
\begin{aligned}
	f^{(3)}_{\mathrm{BFKL}}(\uh,\vh)=&\,16g^2\int\limits_1^{\uh}d\xi_1\int\limits_1^{\vh}d\xi_2\,f^{(4)}_{\mathrm{SF}}(\xi_1,\xi_2)+\frac{\pi}{2}(\nu(\vh)-4g\vh)\\
	&+\frac{\pi}{2}(\nu(\uh)-4g\uh)-4g\left(-\frac{\pi}{4}+\frac{\pi^2}{16}\right).
\end{aligned}
	\label{eq:full_res_f3}
\end{equation}
Putting all contributions together, we obtain the full measure for $\uh>1$ as
\begin{align}
	\log\mu(\uh)=\frac{\sqrt{\lambda}}{2\pi}\Bigg[-4\int\limits_{\theta}^\infty\int\limits_{\theta}^\infty \frac{d\theta_1 d\theta_2}{\sinh(2\theta_1)\sinh(2\theta_2)}\frac{1}{\cosh(\theta_1-\theta_2)}+\pi\left(\frac{\nu(\theta)}{2g}-2\uh\right)+2\pi-\frac{\pi^2}{2}\Bigg],
%	\log\mu(\uh)=&32g^2\int\limits_{1}^{\uh}\int\limits_{1}^{\uh}d\xi_1d\xi_2f^{(4)}_{\mathrm{SF}}(\xi_1,\xi_2)-32g^2\int\limits_{1}^{\uh}\int\limits_{1}^{\uh}d\xi_1d\xi_2f^{(3)}_{\mathrm{SF}}(\xi_1,\xi_2)\nonumber\\
%	&\, +\pi\left(\nu(\uh)-4g\uh\right)+\frac{\sqrt{\lambda}}{2\pi}\left(2\pi-\frac{\pi^2}{2}\right),
	\label{eq:full_measure_ug1}
\end{align}
where we substituted $\uh=\coth(2\theta)$.
The region $\uh<-1$ is obtained by noting that the measure at strong coupling is symmetric under $\uh\leftrightarrow-\uh$.
%%%%%%%%%%%%%%%%%%%%%%%%%

\subsection{$\hat{u},\hat{v}<1$}

We now turn to the other region, where $\hat{u}, \hat{v}<1$.
This region is relevant for the analysis in the main text, as the saddle point turns out to be close to $u_0\approx 0$.
This region is not considered in \cite{Belitsky:2015qla}.
We therefore have to perform the analogous calculations of the quantities appearing in eqs.~(\ref{eq:int_f3}, \ref{eq:def_int_f4}).
We find that the function $\widetilde{\gamma}^{\mathrm{f}}_{-,\uh}(\tau)$ takes a slightly different form and is given by
\begin{equation}
	\widetilde{\gamma}^{\mathrm{f}}_{-,\uh}(\tau)=\frac{\tau}{4g}\left[-\frac{1}{4\sqrt{2}}\sqrtuhml W^+(\tau,\uh)+\frac{1}{4\sqrt{2}}\sqrtuhpl W^+(\tau, -\uh)\right].
	\label{eq:lo_gamma_f3_g}
\end{equation}
The $\tau$-integral can be carried out as in eq.~(\ref{eq:fouriercos}) but due to the assumption that $\uh, \vh <1$ now all integrals are principal value integrals.
Partitioning principal values as
\begin{equation}
	\frac{\mathcal{P}}{x-a}\frac{\mathcal{P}}{x-b}=\frac{\mathcal{P}}{a-b}\left(\frac{\mathcal{P}}{x-a}-\frac{\mathcal{P}}{x-b}\right)+\pi^2\delta(a-b)\delta(x-a)
	\label{eq:part_denom}
\end{equation}
and performing the integrals using the identity
\begin{equation}
	\int\limits_{-1}^1\frac{dk}{\pi}\left(\frac{1+k}{1-k}\right)^{\frac{1}{4}}\frac{\mathcal{P}}{k-p}=-\left(\frac{1+p}{1-p}\right)^{\frac{1}{4}}+\sqrt{2},
\end{equation}
where $|p|<1$, we obtain the result
\begin{equation}
\begin{aligned}
	f^{(3)}_{\mathrm{SF}}(\uh,\vh)=-\frac{1}{64g}\Bigg[&\frac{1}{\uh+\vh}\left(\sqrtuhml\sqrtvhml-\sqrtuhpl\sqrtvhpl\right)\\
	&-\frac{1}{\uh-\vh}\left(\sqrtuhml\sqrtvhpl-\sqrtuhpl\sqrtvhml\right)\Bigg]\\
	&+\frac{\pi}{32g}\delta(\uh-\vh)-\frac{\pi}{32g}\delta(\uh+\vh).
\end{aligned}
	\label{eq:res_f3_sf_l}
\end{equation}

To integrate to the BFKL case, we again use relation eq.~(\ref{eq:bc_kappa_const}) and find
\begin{align}
	f^{(4)}_{\mathrm{BFKL}}(\uh,\vh)&=2\left(-\int\limits_{\uh}^1 d\xi_1\frac{d\kappa^{\mathrm{BFKL}}}{d\xi_1}+\kappa^{\mathrm{BFKL}}(1)\right)\cdot\mathbb{Q}\cdot\mathbb{M}\cdot\left(-\int\limits_{\uh}^1 d\xi_2\frac{d\kappa^{\mathrm{BFKL}}}{d\xi_2}+\kappa^{\mathrm{BFKL}}(1)\right)\nonumber\\
	&=16g^2\int\limits_{\uh}^1d\xi_1\int\limits_{\vh}^1d\xi_2 \,f^{(3)}_{\mathrm{SF}}(\xi_1,\xi_2)+\frac{\sqrt{\lambda}}{2\pi}\left(-\frac{\pi}{2}+\frac{4}{\pi}\right).
\end{align}
Using our result eq.~(\ref{eq:res_f3_sf_l}), substituting $\xi_i=\tanh(2\theta_i)$ and taking the limit $\vh\rightarrow\uh$ we obtain
\begin{equation}
\begin{aligned}
	f^{(4)}_\mathrm{BFKL}(\uh)=\frac{\sqrt{\lambda}}{2\pi}\Bigg[&-\frac{1}{2}\int\limits_{\theta}^\infty\int\limits_\theta^\infty \frac{d\theta_1d\theta_2}{\cosh(2\theta_1)\cosh(2\theta_2)}\left(\frac{1}{\cosh(\theta_1-\theta_2)}-\frac{1}{\cosh(\theta_1+\theta_2)}\right)\\
	&-\frac{\pi}{4}\uh-\frac{\pi}{4}+\frac{4}{\pi}\Bigg],
\end{aligned}
	\label{eq:fin_res_f4}
\end{equation}
where $\theta=\frac{1}{2}\tanh^{-1}(\uh)$.\par
%%%%%%%%%%%%%%%%%
In the same way, we obtain the result for $f^{(3)}_{\mathrm{BFKL}}(\uh)$.
We begin with the modifications for the small fermion function $f^{(4)}_{\mathrm{SF}}(\uh,\vh)$, where the only change from eq.~(\ref{eq:def_int_f4}) is in $\gamma^{\mathrm{f}}_{+,\uh}$, which now reads
\begin{equation}
	\gamma^{\mathrm{f}}_{+,\uh}(\tau)=\frac{\tau}{4g}\left[\frac{1}{4\sqrt{2}}\sqrtuhml W^{-}(\tau,\uh)+\frac{1}{4\sqrt{2}}\sqrtuhpl W^{-}(\tau,-\uh)\right].
	\label{eq:lo_gamma_f4_l}
\end{equation}
Going through the same steps as before we obtain the result
\begin{equation}
\begin{aligned}
	f^{(4)}_{\mathrm{SF}(\uh,\vh)}=-\frac{1}{64g}\Bigg[&\frac{1}{\uh+\vh}\left(\sqrtuhml\sqrtvhml-\sqrtuhpl\sqrtvhpl\right)\\
	&+\frac{1}{\uh-\vh}\left(\sqrtuhml\sqrtvhpl-\sqrtuhpl\sqrtvhml\right)\Bigg]\\
	&-\frac{\pi}{32g}\delta(\uh+\vh)-\frac{\pi}{32g}\delta(\uh-\vh).
\end{aligned}
	\label{eq:fin_res_f4_l}
\end{equation}
\begin{figure}[t]
	\centering
	\includegraphics[scale=1]{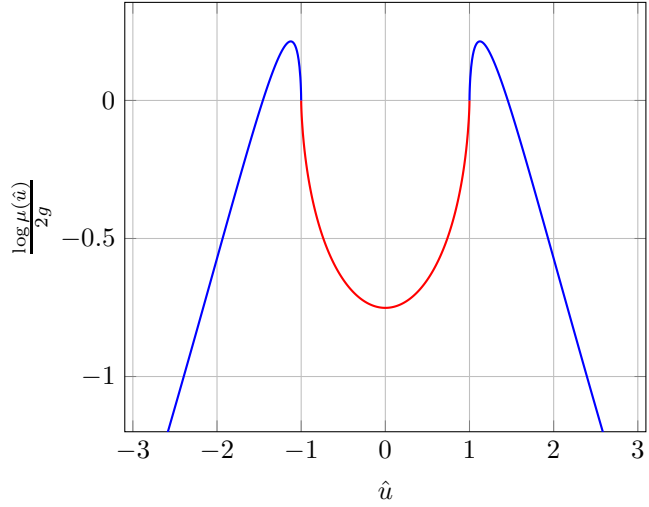}
	\caption{Plot of the strong coupling BFKL measure $\log\mu(\hat{u})/(2g)$. We switch colors from red to blue when $|\hat{u}|$ becomes bigger than one, i.e.\ when the description switches between eq.~(\ref{eq:fin_res_measure_l}) and eq.~(\ref{eq:full_measure_ug1}).}
	\label{fig:plt_measure}
\end{figure}
This can be integrated to the BFKL case using the boundary condition $\widetilde{\kappa}^{\mathrm{BFKL}}(0)=0$ to find
\begin{equation}
\begin{aligned}
	f^{(4)}_\mathrm{BFKL}(\uh,\vh)=&\,2\, \widetilde{\kappa}^{\mathrm{BFKL}}(\uh)\cdot\mathbb{Q}\cdot\mathbb{M}\cdot\widetilde{\kappa}^{\mathrm{BFKL}}(\vh)\\
	=&\,2\left(\int\limits_0^{\uh}d\xi_1\frac{d\widetilde{\kappa}^{\mathrm{BFKL}}}{d\xi_1}\right)\cdot\mathbb{Q}\cdot\mathbb{M}\cdot\left(\int\limits_0^{\vh}d\xi_1\frac{d\widetilde{\kappa}^{\mathrm{BFKL}}}{d\xi_1}\right)
\end{aligned}
\end{equation}
Again, we substitute $\xi_i=\tanh(2\theta_i)$ and take the limit $\vh\rightarrow\uh$ to obtain
\begin{equation}
	f^{(3)}_{\mathrm{BFKL}}(\uh)=\frac{\sqrt{\lambda}}{2\pi}\left[\frac{1}{2}\int\limits_0^\theta\int\limits_0^\theta \frac{d\theta_1d\theta_2}{\cosh(2\theta_1)\cosh(2\theta_2)}\left(\frac{1}{\cosh(\theta_1-\theta_2)}+\frac{1}{\cosh(\theta_1+\theta_2)}\right)-\frac{\pi}{4}\uh\right].
	\label{}
\end{equation}
Finally we can put all results together and find that the BFKL measure in this region to leading order at strong coupling is given by
\begin{align}
	\log\mu^{\mathrm{BFKL}}(\uh)&=\frac{\sqrt{\lambda}}{2\pi}\Bigg[\frac{\pi}{4}\left(2-\pi\right)+ \int\limits_0^\theta\int\limits_0^\theta \frac{d\theta_1d\theta_2}{\cosh(2\theta_1)\cosh(2\theta_2)}\left(\frac{1}{\cosh(\theta_1-\theta_2)}+\frac{1}{\cosh(\theta_1+\theta_2)}\right)\nonumber\\
	&\quad+\int\limits_{\theta}^\infty\int\limits_\theta^\infty \frac{d\theta_1d\theta_2}{\cosh(2\theta_1)\cosh(2\theta_2)}\left(\frac{1}{\cosh(\theta_1-\theta_2)}-\frac{1}{\cosh(\theta_1+\theta_2)}\right)\Bigg]\label{eq:fin_res_measure_l}\\
	&=\frac{\sqrt{\lambda}}{2\pi}\Bigg[\int\limits_{\theta}^\infty\int\limits_\theta^\infty \frac{d\theta_1 d\theta_2}{\cosh(2\theta_1)\cosh(2\theta_2)}\frac{2}{\cosh(\theta_1-\theta_2)}+\frac{\pi}{2}\left(\frac{\nu(\uh)}{g}-2\uh\right)+\pi-\frac{\pi^2}{2}\Bigg]\nonumber,
\end{align}
where we used eq.~(\ref{eq:bfkl_sc}) in the last step. 
This is the formula quoted in the main text, see eq.~(\ref{eq:fin_res_measure_mt}).
We provide a plot of the strong coupling measure in figure \ref{fig:plt_measure}.
\bibliographystyle{JHEP}
\bibliography{tba_subleading}

\end{document}